\documentclass[11pt]{article}
\usepackage[a4paper,margin=1in]{geometry}
\usepackage[T1]{fontenc}
\usepackage[utf8]{inputenc}
\usepackage{lmodern}
\usepackage{microtype}
\usepackage{booktabs}
\usepackage{longtable}
\usepackage{tabularx}
\usepackage{enumitem}
\usepackage[hidelinks]{hyperref}
\usepackage{xurl}
\usepackage{array}
\usepackage{amsmath,amssymb}
\usepackage{titlesec}
\setlist[itemize]{leftmargin=*,itemsep=2pt,topsep=3pt}
\setlist[enumerate]{leftmargin=*,itemsep=2pt,topsep=3pt}
\titleformat{\section}{\large\bfseries}{\thesection}{0.7em}{}
\titleformat{\subsection}{\normalsize\bfseries}{\thesubsection}{0.7em}{}
\newcolumntype{Y}{>{\raggedright\arraybackslash}X}

\title{From Runtime Records to Legal Findings:\\An Evidentiary-Adequacy Criterion for Agentic AI Oversight}
\author{Jeroen Janssen\\Apparens\\\texttt{office@apparens.nl}}
\date{Technical report, July 2026}

\begin{document}
\maketitle

\begin{abstract}
Agentic AI systems generate runtime records, logs, traces, and audit artefacts, but the existence or integrity of such records does not by itself establish that legally operative oversight findings can be recovered from them. This technical report defines an evidentiary-adequacy criterion for a bounded class of determinations: binary findings of fact about specific events and their relations, such as whether protected data crossed a boundary, whether a human could intervene, whether an information barrier held, or whether delegated authority was valid at the moment of use. The criterion states that a runtime record can answer such a determination only if it carries both a typing that maps recorded events to the legally operative category and the relation, such as provenance, authority, derivation, or temporal validity, on which the determination's truth depends. The claim is one of necessity, not sufficiency. It is shown by construction across the defined determination class, with minimality and bounded adequacy stated explicitly. The report instantiates the criterion against selected EU AI Act oversight obligations and explains why tamper-evident logs, generic process frameworks, and provenance structures alone cannot establish the relevant findings. It further relates the argument to requisite variety, the Good Regulator Theorem, and the trace-versus-hyperproperty boundary of runtime verification. Companion materials and the experiment protocol are archived on Zenodo.
\end{abstract}


\section{Introduction}
Agentic AI systems select execution paths at run time. They use models, tools, memory, policies, and harnesses to compose sequences of actions that are not fully enumerated at design time. Enterprise governance has often responded by extending static instruments: point-in-time audits, conformity checklists, periodic review cycles, tamper-evident logs, and retrospective assurance reports. These instruments can be useful, but they do not answer a prior question: under what conditions can a record be read as evidence for a legally operative finding of fact?

This report defines an evidentiary-adequacy criterion for runtime oversight. For a class of determinations that are findings of fact about specific events and relations, a runtime record can answer the determination only if the record carries two features: a typing that maps recorded events to the legally operative category, and a relation, typically provenance, derivation, authority, or temporal validity, on which the determination's truth depends. The criterion is a necessity claim. It is not a claim that any such representation is sufficient for good oversight, compliance, truthfulness, or governance quality.

The worked legal instance is EU AI Act oversight, especially the relation between logging and human oversight obligations. The paper's legal claim is conditional: a record that is not semantically and relationally interpretable cannot serve as evidence that effective oversight was possible for determinations whose truth is not a function of the record. The report does not claim that the EU AI Act mandates a particular ontology or schema.

\subsection{Contribution}
This report makes four contributions. First, it states a determination-answerability criterion over runtime evidence representations. Second, it demonstrates necessity by construction for four in-scope determinations and states minimality and bounded adequacy. Third, it positions the criterion against existing governance, provenance, logging, and runtime-monitoring approaches. Fourth, it gives a pre-registered empirical protocol intended to test whether expert panels can resolve the defined determinations from bare logs, provenance-only records, or typed relational representations.

\subsection{Scope and non-claims}
The report is deliberately narrow. It does not claim novelty for the cybernetic inversion that applies model-dependence to the overseer rather than the AI system. It does not claim that a particular ontology is required. It does not claim that the criterion is sufficient for compliance. It does not claim that requisite variety and monitorability are one theorem. These results are used as convergent support, not as the source of the criterion.

\section{The Evidentiary-Adequacy Criterion}
Let a determination $D$ be a legally operative finding of fact that a supervisory system must be able to establish about an agentic system's behaviour. Let an evidence representation $E$ be the record from which the supervisory system attempts to establish $D$. Let an answering procedure be a decision procedure $f$ that takes $E$ and returns a truth value for $D$.

\textbf{Definition 1 (answerability).} $E$ answers $D$ when there exists a decision procedure $f$ such that $f(E)$ recovers the truth value of $D$, soundly, for the cases in the determination's scope. $E$ fails to answer $D$ when no such procedure exists over $E$, that is, when the truth value of $D$ is not a function of the content of $E$.

\textbf{Selection rule.} A determination is in scope when it is a binary finding of fact about specific events and their relations: a yes-or-no question whose truth value is fixed by what specific events occurred, what authority they ran under, and how they relate to one another. A determination is out of scope when it is an evaluative or statistical property over a distribution of behaviours rather than a finding about particular events. Fairness, robustness, non-discrimination, explainability, and transparency are therefore out of scope for this criterion as stated.

\textbf{Criterion 1 (evidentiary adequacy).} Let $D$ be an in-scope determination of the form: did event-type $X$, of legal category $C$, stand in relation $R$ to event-type $Y$? Then $E$ answers $D$ only if $E$ represents, explicitly or in functionally equivalent form, both:
\begin{enumerate}
  \item a typing that maps the recorded events to the legal category $C$; and
  \item the relation $R$, such as provenance, derivation, authority, or temporal validity, on which the truth of $D$ depends.
\end{enumerate}
A representation carrying neither the typing nor the relation can raise a suspicion, but it cannot establish the finding.

\subsection{Necessity by construction}
Table~\ref{tab:necessity} states the construction. A bare action log is assumed to record that an action occurred, when it occurred, and whether it was permitted, without legal typing of data, channels, or actors, and without provenance, derivation, authority state, or temporal validity.

\begin{table}[ht]
\centering
\small
\begin{tabularx}{\textwidth}{p{0.23\textwidth}Y Y Y}
\toprule
\textbf{Determination} & \textbf{Truth depends on} & \textbf{Why a bare log fails} & \textbf{Minimal addition} \\
\midrule
Did protected data leave the controlled environment? & Whether protected-class data crossed an external boundary. & The log may show that a send occurred, but not the protected class of the content or the boundary type. & Data-category typing and channel/boundary typing. \\
Was a human able to intervene? & Whether a human held authority and a usable intervention window at the decision point. & The log shows execution, not reachable override authority or time-to-intervene. & Authority-state and intervention-window relation. \\
Did the information barrier hold? & Whether information derived from a high-side source reached a low-side sink. & The log records actions, not derivation between source and sink. & Provenance linking source read to sink send, plus side typing. \\
Was delegated authority valid when used? & Whether authority was granted, unrevoked, in scope, and valid at the instant of use. & The log records use, not the temporal validity of the authority state. & Delegation, revocation, scope, and temporal binding. \\
\bottomrule
\end{tabularx}
\caption{Necessity by construction: why a bare log cannot answer each determination.}
\label{tab:necessity}
\end{table}

In each row, the truth value of $D$ depends on information that the bare log does not contain. The determination is therefore not a function of the log's content. By Definition 1, the bare log fails to answer it.

\subsection{Minimality and bounded adequacy}
The two-feature set is minimal. If typing is dropped, a record can show a boundary crossing without showing that the content was of a protected legal category. If the relation is dropped, a record can type two events without showing whether one derived from the other or whether authority was valid when used. Neither feature is redundant.

The claim is also bounded. The relation feature must be read broadly to include provenance, derivation, authority, and temporal structure. Under that reading, the set is adequate for the four determinations and for determinations of the same form. The report does not claim adequacy for all properties that a regulator, auditor, court, or board might wish to establish.

\subsection{Why sufficiency fails}
A record satisfying the criterion can still fail to support effective oversight. It may be incomplete, curated, falsely typed, produced in bad faith, or paired with incompetent decision-making or delayed intervention. The criterion buys legibility, not truthfulness. It says when a record can be read for a determination; it does not make the record honest.

\section{Existing Frameworks Against the Criterion}
The criterion distinguishes producing records from producing findings. Process frameworks govern organisational functions. Logging frameworks produce event records. Provenance frameworks carry derivation. None, by itself, necessarily carries both the legal typing and the relation required for the determinations in scope.

\begin{table}[ht]
\centering
\small
\begin{tabularx}{\textwidth}{p{0.24\textwidth}p{0.22\textwidth}Y}
\toprule
\textbf{Framework or mechanism} & \textbf{Record/finding posture} & \textbf{Gap against the criterion} \\
\midrule
NIST AI Risk Management Framework & Process and outcomes framework. & It does not specify a runtime evidentiary representation that types events and carries relations. \\
ISO/IEC 42001 & Management-system standard. & It can certify or assess management-system operation, but does not define the content of a record from which the legal finding is recoverable. \\
EU AI Act Article 12 logging & Automatic event recording for traceability. & The logging duty does not, on its face, require that logged events be typed to legal categories and connected by relations. \\
Runtime verification and enforcement & Mediates actions and may produce enriched traces. & It can answer more when instrumented, but a black-box single-path monitor cannot evidence relational properties without system knowledge. \\
Provenance models & Derivation relations. & They carry relation, but not the legal typing needed to answer legally operative determinations. \\
\bottomrule
\end{tabularx}
\caption{Existing frameworks and mechanisms against the answerability criterion.}
\end{table}

This is not a criticism of those frameworks as such. It marks the boundary between governance process, trace integrity, provenance, and supervisory answerability.

\section{Convergent Supporting Results}
The criterion is argued from the determinations. Three established results support the same structural gap from different directions.

\subsection{Requisite variety}
Ashby's Law of Requisite Variety states that a regulator can hold an essential variable in range across disturbances only if its effective variety can match the variety that must be compensated \cite{ashby1956}. Applied qualitatively, a point-in-time audit acting as the sole oversight mechanism is in the wrong class for a path-selecting agent fleet. It acts once against a system that generates behavioural variety continuously. This is an application, not an information-theoretic proof.

The relevant point for this paper is narrower: a legal condition becomes an essential variable only if it is rendered measurable, sensed through a channel, connected to a decision rule, and tied to an intervention. Where the record cannot be read for the legally operative fact, it cannot serve as the sensing channel of such a regulator.

\subsection{Good Regulator Theorem}
Conant and Ashby argue that every good regulator must be a model of the system it regulates \cite{conant1970}. The report uses this result only under its coupling condition: it applies to a supervisory system that senses, decides, and can affect future states. A purely retrospective evaluator is outside that theorem. The theorem supports model-dependence, not the legal content of the model. The content - typing and relation - comes from the criterion.

The inversion of model-dependence onto the overseer is not claimed as novel. Aguirre develops a related inversion at civilizational scale in the context of superintelligent agents \cite{aguirre2025}. This report applies the issue to enterprise-scale oversight and to legal-evidentiary answerability.

\subsection{Runtime monitorability and hyperproperties}
Schneider characterised enforceable security policies for execution monitors observing single executions \cite{schneider2000}. Clarkson and Schneider distinguish trace properties from hyperproperties, including noninterference as a 2-safety hyperproperty \cite{clarkson2010}. A relational property cannot generally be evidenced from a single black-box trace. Gray-box monitoring can lift the limit when the monitor is given system knowledge \cite{stucki2019}. That condition is the security vocabulary version of the criterion: the record must carry enough typing and relation for the determination to be computed.

\section{Information-Barrier Scenario}
Consider two agents in one organisation. Agent A has access to a deal room containing material non-public information. Agent B has authority to send messages to a trading desk. Each action is individually permitted: A reads the deal room, and B sends to the desk. The possible violation is not either action alone. It is whether information derived from A's high-side read reached B's low-side send.

\begin{table}[ht]
\centering
\small
\begin{tabularx}{\textwidth}{p{0.28\textwidth}Y}
\toprule
\textbf{Layer} & \textbf{Can it answer ``did the barrier hold''?} \\
\midrule
Per-action access policy & No. It sees one actor, one resource, and one action at a time. \\
Single-path runtime monitor & No, if the breach is defined across paths and the monitor lacks relational knowledge. \\
Bare behavioural trace & No. It records actions, not whether information flowed. \\
Typed provenance representation & In principle yes, where the read, send, source, sink, and derivation relation are represented. \\
\bottomrule
\end{tabularx}
\caption{The information-barrier scenario by governance layer.}
\end{table}

If the barrier is formulated as noninterference, the issue is relational across executions. If formulated operationally as taint tracking, it becomes monitorable on a single enriched trace only because the trace carries the derivation relation and the typing of source and sink. Under either formulation, the unenriched action log cannot answer the finding.

The boundary is important. The criterion applies to flows crossing provenance-capturable surfaces. It does not capture model-internal latent state, opaque tool execution, summarisation that destroys provenance, human transfer between screens, or timing side channels unless those channels are instrumented.

\section{EU AI Act Instance}
The EU AI Act instance is stated conditionally. Article 12 concerns automatic recording for high-risk AI systems. Article 14 concerns human oversight, including the ability of natural persons to understand, intervene, and override. Article 26 places monitoring and log-keeping obligations on deployers, and Article 72 concerns post-market monitoring \cite{euai2024}.

A bare log may satisfy part of the recording duty. The report does not claim that Article 12 itself mandates a semantic representation. Article 14 is the stronger anchor. Oversight that cannot answer the operative determination cannot evidence that effective oversight was possible for that determination. Therefore, a record that satisfies the letter of logging can still fail to support oversight if the needed typing and relation are absent.

The sufficiency direction is weak: a structured supervisory record can serve as evidence input toward Article 14 oversight and Article 26 monitoring, but it is not itself a discharge of those duties. The necessity direction is stronger: a record that is not semantically and relationally interpretable cannot serve as evidence for a finding whose truth is not a function of the record.

Legal dates and implementation dependencies are perishable. The enacted high-risk obligation dates and any proposed deferrals should be verified against the Official Journal and the current Commission materials before relying on them in compliance work.

\section{Two Oversight Models and Predictions}
The criterion separates two models of oversight. The current model relies on static or periodic audit plus tamper-evident action records. The semantic model produces a runtime sensing channel that types events and carries relations, so a determination is recovered by a procedure over the record.

\begin{table}[ht]
\centering
\small
\begin{tabularx}{\textwidth}{p{0.25\textwidth}Y Y}
\toprule
\textbf{Property} & \textbf{Current oversight} & \textbf{Semantic oversight} \\
\midrule
Record integrity & Present if hash-chained or timestamped. & Present if integrity controls are retained. \\
Answerability & Often absent for relational findings. & Present for represented, provenance-capturable flows. \\
Interpretability & Manual reconstruction. & Procedure over typed relational record. \\
Relational monitorability & Weak from a single black-box trace. & Stronger with gray-box instrumentation and relation. \\
Article 14 evidence input & Cannot evidence determinations it cannot answer. & Can serve as evidence input, not as complete discharge. \\
\bottomrule
\end{tabularx}
\caption{Two models of runtime oversight.}
\end{table}

Two falsifiable predictions follow. First, organisations relying only on logs without typing and relation will systematically fail to answer the in-scope determinations, while typed relational records will answer them for provenance-capturable flows. Second, increasing logging volume without adding typing and relation will exhibit diminishing oversight value for the remaining unanswered determinations.

\section{Empirical Validation Protocol}
The criterion is testable. The planned experiment is a between-groups design with three expert panels. All groups receive the same set of cases and determinations. Group A receives an action log only: events, timestamps, and permit decisions. Group B receives the action log plus a provenance graph: relation without legal typing. Group C receives the typed relational representation: typing and relation both present.

The primary measure is the proportion of determinations resolved correctly against ground truth. Secondary measures include ``cannot determine'' responses, median resolution time, inter-rater agreement, and confidence. The falsification condition is explicit: if Group A resolves relational determinations at a rate not distinguishable from Group C, the criterion is false or trivial for this setting. If Group B resolves typing-dependent determinations at the rate of Group C, the minimality claim is challenged.

The protocol is fixed in Appendix~\ref{app:protocol}. Results are not reported in this version.

\section{Risks, Limits, and Controls}
The criterion is necessity, not sufficiency. It does not provide a complete theory of oversight. It does not guarantee truthful records. It does not solve capture at uninstrumented surfaces. It does not convert a management standard into legal compliance. It does not replace human legal judgment.

The strongest objections are preserved as tests rather than hidden. The criterion may be trivial if deployed records already carry typing and relation. The cybernetic inversion is not novel and is attributed to prior work. Article 14 does not mandate a particular ontology, and the report does not claim otherwise. The report is strongest where it stays narrow: records without the needed semantic and relational content cannot establish findings whose truth depends on that missing content.

\section{Implications}
For boards and accountable functions, the distinction is between integrity and answerability. A tamper-evident log can show that a record was not altered after the fact and still fail to show what legally operative fact the record establishes. For standardisation, the implication is that logging and oversight obligations need an evidentiary representation layer. For runtime design, the implication is to treat oversight as a loop: sensing, decision, and intervention. The sensing channel must answer the determination before the decision and intervention functions can operate intelligently.

The structure is not specific to AI. Autonomous trading systems, medical devices, critical infrastructure controllers, and distributed software agents can all generate determinations whose truth depends on events and relations. EU AI Act oversight is the worked instance, not the conceptual boundary.

\section{Conclusion}
For a defined class of findings of fact about autonomous-system execution paths, a record is evidence of the finding only when it carries both the typing that maps events to the operative legal category and the relation on which the finding's truth depends. A record lacking those features may raise suspicion, but it cannot establish the finding, however tamper-evident it is. The claim is narrower than an impossibility result about oversight and narrower than a theory of knowledge. It is a criterion for whether a runtime record can answer a legally operative determination at all.

\section*{Data and materials availability}
Companion materials, including the prior working paper record and supporting artefacts, are available on Zenodo: DOI \href{https://doi.org/10.5281/zenodo.21025237}{10.5281/zenodo.21025237}. This report also refers to the supervisory-evidence ontology deposit, DOI \href{https://doi.org/10.5281/zenodo.19758441}{10.5281/zenodo.19758441}.

\section*{AI disclosure}
Generative AI tools were used for adversarial review, editing, formatting, and preparation of this arXiv technical-report edition. The author directed, checked, and approved the claims, citations, and conclusions and remains responsible for the content.

\section*{Licence}
This arXiv submission is made under the arXiv perpetual, non-exclusive license. The underlying working-paper materials are published under CC BY 4.0 as indicated in the Zenodo record.

\appendix
\section{Claim Inventory and Falsification Register}\label{app:claims}
This appendix decomposes the argument into discrete claims, grades them by type and load-bearing status, and states falsification conditions for the most exposed claims. The purpose is adversarial transparency: the criterion is intended to be testable rather than merely asserted.

\subsection{Legend}
Claim type: Def = definitional; Int = interpretive; Inf = inferential; Emp = empirical; Leg = legal; Nov = novelty; Nrm = normative. Load-bearing status: Critical = the thesis collapses or must be materially narrowed if false; Supporting = failure weakens but does not collapse; Peripheral = illustrative. Evidential grade: A = canonical or established; B = well-supported; C = defensible inference, contestable; D = conjecture or proof-of-absence; E = unverified or perishable.

\subsection{Master register}
\small
\begin{longtable}{p{0.08\textwidth}p{0.54\textwidth}p{0.12\textwidth}p{0.12\textwidth}p{0.08\textwidth}}
\toprule
\textbf{ID} & \textbf{Claim summary} & \textbf{Type} & \textbf{Load} & \textbf{Grade} \\
\midrule
\endfirsthead
\toprule
\textbf{ID} & \textbf{Claim summary} & \textbf{Type} & \textbf{Load} & \textbf{Grade} \\
\midrule
\endhead
K1 & Agentic systems select execution paths at run time and those paths can be steered by content read by the agent. & Def/Emp & Critical & A \\
K2 & A determination is answerable from a record iff the determination's truth value is a function of the record's content. & Def & Critical & A \\
K3 & The in-scope class is binary findings of fact about specific events and relations, not evaluative or statistical properties. & Def & Critical & B \\
K4 & An in-scope determination is answerable only if the record carries legal typing and the relation on which truth depends. & Inf & Critical & B \\
K5 & A representation carrying neither typing nor relation can raise suspicion but cannot establish a finding. & Inf & Critical & B \\
K6 & Necessity holds by construction for the four determinations used in the report. & Inf & Critical & B \\
K7 & The set \{typing, relation\} is minimal for the defined class. & Inf & Critical & B \\
K8 & The set is adequate only for the bounded class when relation includes provenance, authority, and temporal structure. & Inf & Critical & C \\
K9 & Sufficiency fails because a legible record can still be captured or false. & Inf & Critical & B \\
K10 & The criterion is satisfied by structure, not by a unique schema. & Int/Inf & Critical & B \\
K11 & No current governance framework, by itself, produces a record satisfying the criterion for the in-scope class. & Emp/Inf & Critical & C \\
K12 & Tamper-evidence establishes integrity, not answerability. & Int/Leg & Supporting & B \\
K13 & Static audit as sole mechanism lacks effective variety for runtime-dependent variables. & Inf & Supporting & C \\
K14 & A legal condition becomes a regulatory variable only when measurable, sensed, decisionable, and intervention-tied. & Def/Inf & Supporting & C \\
K15 & The Good Regulator Theorem supports model-dependence only for coupled regulators, not retrospective evaluators. & Int & Critical & B \\
K16 & The legal content of the model is supplied by the criterion, not by the theorem. & Int & Critical & B \\
K17 & Relational properties such as noninterference require more than a single black-box trace unless instrumentation adds system knowledge. & Emp/Formal & Supporting & A \\
K18 & Schneider's enforceable-monitor class is contained in safety properties. & Emp/Formal & Supporting & A \\
K19 & The cybernetic and monitorability results converge but are not one theorem. & Inf & Supporting & D \\
K20 & The cybernetic inversion onto the overseer is prior work and is not claimed as novel. & Nov/Emp & Critical & B \\
K21 & The contribution is the conjunction of criterion, EU-law instance, gap analysis, and test protocol. & Nov & Critical & C \\
K22 & EU AI Act Article 12 concerns logging; Article 14 concerns effective human oversight. & Leg & Critical & A \\
K23 & A bare log may satisfy Article 12 while failing to support Article 14 oversight. & Leg/Inf & Critical & C \\
K24 & A non-interpretable record cannot evidence that effective oversight was possible for determinations not recoverable from it. & Leg/Inf & Critical & C \\
K25 & Current and proposed EU AI Act dates are perishable and must be checked against primary legal sources. & Leg/Emp & Supporting & B \\
K26 & Prior legal analysis of agents under EU law is compatible with the traceability requirement. & Emp/Cite & Supporting & B \\
K27 & The criterion is bounded to provenance-capturable surfaces. & Emp/Inf & Critical & C \\
K28 & The two oversight models yield falsifiable predictions. & Inf & Supporting & B \\
K29 & The between-groups experiment discriminates the criterion from triviality. & Emp & Critical & C \\
K30 & The supervisory-evidence ontology deposit is an existence proof, not a validated instrument. & Emp & Supporting & A \\
K31 & Cybernetic audit theory treats audit as regulatory/trust function but not as this agentic trace-evidence criterion. & Int/Emp & Supporting & B \\
K32 & The criterion is not a theory of knowledge. & Def & Supporting & B \\
K33 & The structure generalises beyond AI but is demonstrated only for the EU AI Act instance. & Inf & Supporting & C \\
\bottomrule
\end{longtable}
\normalsize

\subsection{Falsification targets}
The at-risk critical claims are K8, K11, K21, K23, K24, K27, and K29. K4, K6, and K7 are falsified by an in-scope determination that a bare, untyped, provenance-free event log answers on its own. K8 is falsified by an in-scope determination that cannot be answered even when typing and the broad relation feature are present. K11 is falsified by a governance framework in current use that, by itself, produces a record typing events to legal categories and carrying the relations on which findings depend. K23 and K24 are falsified by a legal reading under which bare logging evidences effective oversight for these determinations. K27 is falsified by showing that the named uninstrumented channels are provenance-capturable by the proposed representation. K29 is falsified if the action-log-only group resolves relational determinations as well as the typed-relational group.

\section{Experiment Protocol}\label{app:protocol}
This appendix specifies the determination-answerability experiment. It is a protocol, not a results section.

\subsection{Hypothesis}
For determinations that depend on a relation, such as the information barrier and the intervention window, the expected resolution order is Group C $>$ Group B $>$ Group A, with Group A at or near floor. For a determination that depends on typing but not relation, Group C is expected to outperform both Group A and Group B, isolating the typing feature.

\subsection{Design}
The experiment uses a between-groups design with three independent expert panels, one condition each, to avoid learning effects. Participants are practitioners with expertise in data protection, AI assurance, audit, or related governance functions. Panel size is to be fixed by power analysis before recruitment.

\subsection{Materials and conditions}
The case basis is the EU automated-decision-making enforcement sample associated with the supervisory-evidence deposit. Each case is rendered in three record conditions:
\begin{itemize}
  \item \textbf{Condition A:} action log only: events, timestamps, and permit decisions.
  \item \textbf{Condition B:} action log plus provenance graph: relation present, legal typing absent.
  \item \textbf{Condition C:} typed relational representation: typing and relation both present.
\end{itemize}
Each case presents the same determinations: protected boundary crossing, human intervention possibility, information-barrier status, and delegated-authority validity.

\subsection{Ground truth}
Each case's determinations are coded to ground truth by two independent coders. Disagreements are adjudicated. Cohen's kappa is reported for the case coding. This ground-truth procedure is separate from any single-coder limitations in the earlier source materials.

\subsection{Procedure}
Each participant receives only the cases in the assigned condition. For each determination, the participant records one of three responses: yes, no, or cannot determine from this record. The last response is explicit and encouraged because answerability failure should not be conflated with guessing. Participants also record time to resolve and confidence on an ordinal scale.

\subsection{Measures}
Primary measure: proportion of determinations resolved correctly against ground truth, by condition and determination type. Secondary measures: proportion answered ``cannot determine,'' median time to resolve, within-panel agreement using Fleiss kappa, and confidence.

\subsection{Falsification}
The criterion is falsified or shown trivial if Group A resolves relational determinations correctly at a rate not distinguishable from Group C. The minimality claim is challenged if Group B resolves typing-dependent determinations at the rate of Group C. The ``cannot determine'' rate is analysed as a manifest check that poor performance in Group A reflects answerability failure rather than participant incompetence.

\subsection{Status}
Design and materials are fixed in protocol form. Recruitment, power analysis, and execution are pending. The report does not presuppose the result.


\begin{thebibliography}{99}
\bibitem{aguirre2025} Aguirre, A. (2025). \emph{Control Inversion: Why the Superintelligent AI Agents We Are Racing to Create Would Absorb Power, Not Grant It}. Future of Life Institute. \url{https://control-inversion.ai/}
\bibitem{anderson1972} Anderson, J. P. (1972). \emph{Computer Security Technology Planning Study}. ESD-TR-73-51. United States Air Force Electronic Systems Division.
\bibitem{argyris1978} Argyris, C. and Sch\"on, D. A. (1978). \emph{Organizational Learning: A Theory of Action Perspective}. Addison-Wesley.
\bibitem{ashby1956} Ashby, W. R. (1956). \emph{An Introduction to Cybernetics}. Chapman \& Hall.
\bibitem{beer1979} Beer, S. (1979). \emph{The Heart of Enterprise}. John Wiley \& Sons.
\bibitem{chiodo2026} Chiodo, M., M\"uller, D., Siewert, P., Wetherall, J.-L., Yasmine, Z. and Burden, J. (2026). Formalising Human-in-the-Loop: Computational Reductions, Failure Modes, and Legal-Moral Responsibility. Preprint.
\bibitem{clarkson2010} Clarkson, M. R. and Schneider, F. B. (2010). Hyperproperties. \emph{Journal of Computer Security}, 18(6), 1157--1210.
\bibitem{conant1970} Conant, R. C. and Ashby, W. R. (1970). Every good regulator of a system must be a model of that system. \emph{International Journal of Systems Science}, 1(2), 89--97.
\bibitem{espejo2001} Espejo, R. (2001). Auditing as a trust creation process. \emph{Systemic Practice and Action Research}, 14(2), 215--236.
\bibitem{euai2024} European Commission (2024). Regulation (EU) 2024/1689 of the European Parliament and of the Council laying down harmonised rules on artificial intelligence (Artificial Intelligence Act). OJ L, 2024/1689.
\bibitem{commission2026} European Commission (2026). An agile Digital Rulebook for the EU and Digital Omnibus on AI. Shaping Europe's Digital Future. \url{https://digital-strategy.ec.europa.eu/en/policies/digital-rulebook}
\bibitem{janssen2026a} Janssen, J. (2026a). From Battlefield to Boardroom: Strategic Red Teaming as an Epistemic Governance Instrument in the Age of AI. Working paper. SSRN. \url{https://papers.ssrn.com/sol3/papers.cfm?abstract_id=6860020}
\bibitem{janssen2026b} Janssen, J. (2026b). A Supervisory-Evidence Ontology for Agentic AI under EU Law: Candidate Minimum Conceptual Set and Temporal Extension. Working paper. Zenodo. DOI: \href{https://doi.org/10.5281/zenodo.19758441}{10.5281/zenodo.19758441}.
\bibitem{janssen2026c} Janssen, J. (2026c). From Record to Finding: Why Tamper-Proof Logs Cannot Establish Legal Oversight of Agentic AI. Working paper. Zenodo. DOI: \href{https://doi.org/10.5281/zenodo.21025237}{10.5281/zenodo.21025237}.
\bibitem{nannini2026} Nannini, L., Leon Smith, A., Maggini, M. J., Panai, E., Feliciano, S., Tiulkanov, A., Maran, E., Gealy, J. and Bisconti, P. (2026). AI Agents Under EU Law: A Compliance Architecture for AI Providers. Preprint, arXiv:2604.04604.
\bibitem{power2007} Power, M. (2007). \emph{Organized Uncertainty: Designing a World of Risk Management}. Oxford University Press.
\bibitem{saltzer1975} Saltzer, J. H. and Schroeder, M. D. (1975). The protection of information in computer systems. \emph{Proceedings of the IEEE}, 63(9), 1278--1308.
\bibitem{schneider2000} Schneider, F. B. (2000). Enforceable security policies. \emph{ACM Transactions on Information and System Security}, 3(1), 30--50.
\bibitem{stucki2019} Stucki, S., S\'anchez, C., Schneider, G. and Bonakdarpour, B. (2019). Gray-box monitoring of hyperproperties. In \emph{Formal Methods - The Next 30 Years}, LNCS 11800, 406--424. DOI: 10.1007/978-3-030-30942-8\_25.
\bibitem{thobani2024} Thobani, I. (2024). A triviality worry for the internal model principle. \emph{Synthese}, 204(1), article 36. DOI: 10.1007/s11229-024-04693-x.
\bibitem{touchette2000} Touchette, H. and Lloyd, S. (2000). Information-theoretic limits of control. \emph{Physical Review Letters}, 84(6), 1156--1159.
\bibitem{virgo2025} Virgo, N., Biehl, M., Baltieri, M. and Capucci, M. (2025). A ``Good Regulator Theorem'' for embodied agents. Preprint, arXiv:2508.06326.
\bibitem{wang2025a} Wang, C. L., Singhal, T., Kelkar, A. and Tuo, J. (2025). MI9: An integrated runtime governance framework for agentic AI. Preprint, arXiv:2508.03858.
\bibitem{wang2025b} Wang, H., Poskitt, C. M. and Sun, J. (2025). AgentSpec: Customizable runtime enforcement for safe and reliable LLM agents. Preprint, arXiv:2503.18666.
\end{thebibliography}
\end{document}